\documentclass[12pt,preprint,url]{aastex}
\usepackage{amsbsy}
\usepackage{amsmath}
\newcommand{\be}{\begin{equation}}
\newcommand{\ee}{\end{equation}}

\def\ltsima{$\; \buildrel < \over \sim \;$}

\def\lsim{\lower.5ex\hbox{\ltsima}}
\def\gtsima{$\; \buildrel > \over \sim \;$}
\def\gsim{\lower.5ex\hbox{\gtsima}}
\shorttitle{Glitch size-waiting-time correlations}
\shortauthors{Melatos et al.}

\begin{document}
\title{Size-waiting-time correlations in pulsar glitches}

\author{A. Melatos\altaffilmark{1,2} 
 and G. Howitt \altaffilmark{1,2}
 and W. Fulgenzi \altaffilmark{1}}

\email{amelatos@unimelb.edu.au}

\altaffiltext{1}{School of Physics, University of Melbourne,
 Parkville, VIC 3010, Australia}

\altaffiltext{2}{Australian Research Council Centre of Excellence
 for Gravitational Wave Discovery (OzGrav)}

\begin{abstract}
\noindent 
Few statistically compelling correlations are found
in pulsar timing data
between the size of a rotational glitch and the time to the preceding glitch
(backward waiting time) or the succeeding glitch (forward waiting time),
except for a strong correlation between sizes and forward waiting times
in PSR J0537$-$6910.
This situation is counterintuitive,
if glitches are threshold-triggered events, 
as in standard theories
(e.g.\ starquakes, superfluid vortex avalanches).
Here it is shown that the lack of correlation emerges naturally,
when a threshold trigger is combined with secular stellar braking
slower than a critical, calculable rate.
The Pearson and Spearman correlation coefficients are computed
and interpreted within the framework
of a state-dependent Poisson process.
Specific, falsifiable predictions are made regarding what objects
currently targeted by long-term timing campaigns should develop strong
size-waiting-time correlations, as more data are collected in the future.
\end{abstract}


\section{Introduction 
 \label{sec:cor1}}
Glitches are impulsive, erratically occurring, spin-up events
which interrupt the secular, electromagnetic spin down of a 
rotation-powered pulsar.
As the number of recorded events rises,
\footnote{
Electronic access to up-to-date glitch catalogues is available at the
following locations on the World Wide Web:
{\tt http://www.jb.man.ac.uk/pulsar/glitches/gTable.html} 
(Jodrell Bank Centre for Astrophysics)
and
{\tt http://www.atnf.csiro.au/people/pulsar/psrcat/glitchTbl.html}
(Australia Telescope National Facility).
\label{foot:cor1}
}
there is growing evidence that glitching pulsars divide into two classes:
Poisson-like glitchers,
whose waiting times and sizes are described by exponential and power-law
probability density functions (PDFs) respectively;
and quasiperiodic glitchers,
whose waiting times and sizes are distributed roughly normally
around characteristic values
\citep{won01,mel08,esp11,onu16,yu17,how18}.
The physical mechanism that triggers glitch activity remains a mystery;
see \citet{has15} for a recent review.
Broadly speaking, however, it is believed that electromagnetic braking
increases the elastic stress and differential rotation in the star,
which then relax abruptly via some combination of starquakes and
superfluid vortex avalanches, when a threshold is exceeded
\citep{and03,mid06,gla09,chu10b,war11}.

Intuitively one expects sizes and waiting times to correlate strongly
in a threshold-driven, stress-release process,
where `stress' refers to any disequilibrium variable
including differential rotation.
For example, after a larger glitch, one expects a longer delay until
the next glitch, while the stress reservoir is replenished.
That is, there should be a strong positive correlation between
sizes and {\em forward} waiting times.
Conversely, after a longer waiting time, 
one expects a larger glitch,
because the stress reservoir is fuller.
That is, there should also be a strong positive correlation between
sizes and {\em backward} waiting times.
The above intuition rests implicitly on the assumption,
that the stress reservoir is mostly emptied by each relaxation event.

In contrast, size-waiting-time correlations are rare 
in pulsar glitch data.
A strong, linear correlation of $6.5\,{\rm days}\,\mu{\rm Hz}^{-1}$
is observed between sizes and forward waiting times in PSR J0537$-$6910
\citep{mid06,fer18,ant18},
which can be exploited to predict reliably the epoch of the next glitch;
see the `staircase plot' in Fig.\ 8 in \citet{mid06}.
A similar claim has been made regarding PSR J1645$-$0317,
where the slope of the correlation is measured to be
$0.38\,{\rm days}\,{\rm pHz}^{-1}$
\citep{sha09}.
However the glitches in PSR J1645$-$0317 rise gradually over $\sim 1 \,{\rm yr}$
and do not belong to the class of impulsive events studied in this paper.
Beyond these two examples,
there is scant evidence to date for statistically compelling correlations 
between sizes and forward waiting times
\citep{yua10}.
Moreover there is no evidence at all for a statistically significant correlation
between sizes and backward waiting times in any object
\citep{yua10,ful17}
nor in microglitches
\citep{onu16}.
Pulsar glitches are not unique in this regard.
The absence of size-waiting-time correlations is mirrored in many
threshold-driven, stick-slip, stress-release systems in nature,
including sandpiles, earthquakes, solar flares,
and flux tube avalanches in type II superconductors
\citep{lu91,fie95,jen98,whe00b,sor04}.
In these self-organized critical systems,
only a small fraction of the stress reservoir empties at 
each relaxation event.
The process randomly releases historical stress
accumulated over an extended period covering many events, 
so that the stress released by any individual event 
is sometimes less than and sometimes greater than
the stress added since the previous event
\citep{jen98,mel08}.
Incomplete reservoir depletion is also inferred in some
quasiperiodic glitchers,
e.g.\ PSR J0537$-$6910
\citep{ant18}.

In this paper, we show quantitatively that sufficiently rapid
electromagnetic braking produces a size-waiting-time correlation
in certain pulsars,
even when the microscopic dynamics
of the underlying, self-organized critical process are uncorrelated,
e.g.\ as in superfluid vortex avalanches
\citep{war08,mel09,war11}.
The paper is structured as follows.
In \S\ref{sec:cor2},
we review the statistical evidence for size-waiting-time correlations 
in the seven pulsars with the largest glitch samples,
using the latest data from the Jodrell Bank Centre for Astrophysics
and Australia Telescope National Facility catalogues 
(see footnote \ref{foot:cor1})
\citep{man05,esp11}.
In \S\ref{sec:cor3} and \S\ref{sec:cor4}, we interpret the data 
in terms of a state-dependent Poisson process
\citep{cox55,dal07,whe08,ful17}
and show that there exists a critical spin-down rate,
above which a strong correlation emerges between sizes and
forward waiting times.
The theory is quantitative and predictive and does not depend
on the microphysics of the glitch trigger.
We close in \S\ref{sec:cor5}
by presenting a ranked list of targets predicted to display
strong correlations,
as a guide to designing the next generation of glitch monitoring campaigns
at radio wavelengths,
e.g.\ with phased arrays like LOFAR
\citep{kra10},
UTMOST \citep{cal16},
and the Square Kilometer Array,
as well as at other wavelengths, e.g.\ gamma rays
\citep{ray11,cla17}.

We emphasize that the state-dependent Poisson process
analysed here and by \citet{ful17} is a meta-model
which is agnostic about the glitch microphysics.
It applies to any threshold-based trigger mechanism,
where the star is driven slowly away from equilibrium by
electromagnetic spin down and releases the cumulative stress impulsively,
e.g.\ via superfluid vortex avalanches (differential rotation)
or starquakes (elastic stresses).
In this paper, we extend the framework in \citet{ful17}
by developing size-waiting-time correlations as a new, quantitative,
observational test of the model.
Specifically, 
we apply the framework to existing glitch catalogues for the first time
(\S\ref{sec:cor2}),
calculate correlation coefficients theoretically
as functions of the spin-down rate and other variables
(\S\ref{sec:cor4}, Appendix \ref{sec:corappa}),
present a new recipe for inverting the correlation data to infer
nuclear parameters, e.g.\ pinning strength
(\S\ref{sec:cor4}, \S\ref{sec:cor5}),
and identify specific objects as targets for future correlation studies
(\S\ref{sec:cor5}).

\section{Data
 \label{sec:cor2}}
\begin{figure}
\begin{center}
\includegraphics[width=13cm,angle=0]{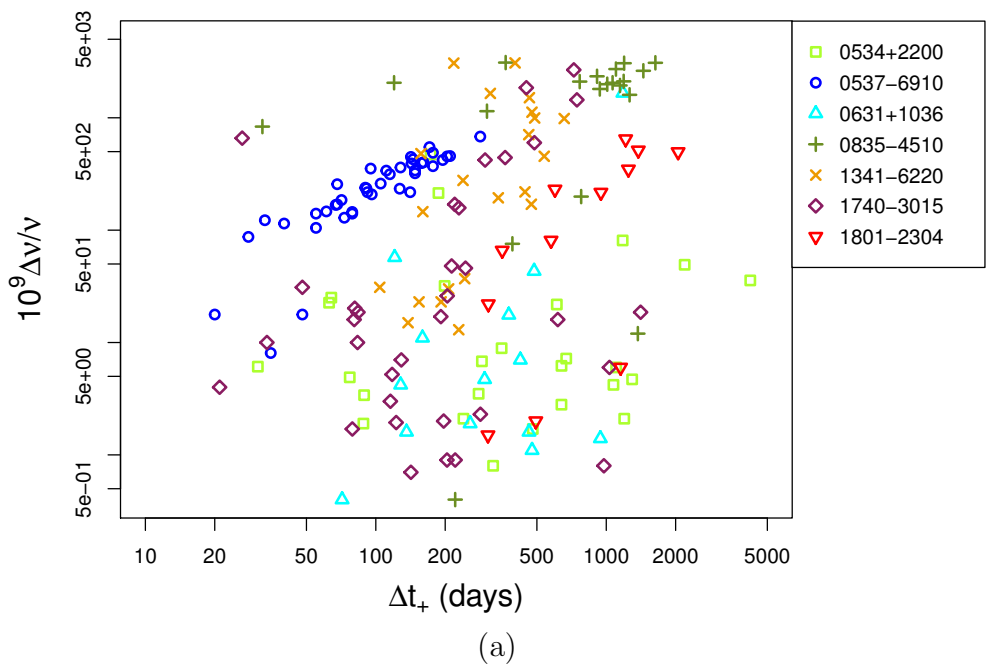}
\\
(a)
\\
\vspace*{0.5cm}
\includegraphics[width=13cm,angle=0]{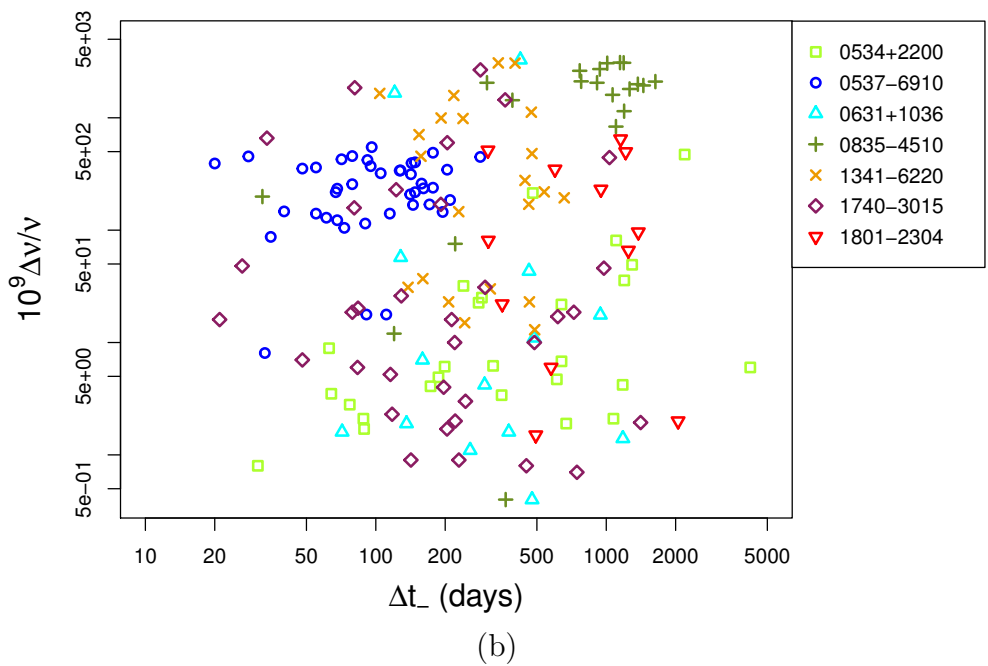}
\\
(b)
\end{center}
\caption{
Log-log plot of fractional size $s$ (multiplied by $10^9$) versus 
(a) forward waiting time $\Delta t_+$ (in days) and
(b) backward waiting time $\Delta t_-$ (in days)
for the seven most active glitching pulsars in Table \ref{tab:cor1}.
}
\label{fig:cor1}
\end{figure}

Advances in pulsar timing methods,
including multibeam surveys and multifrequency ephemerides,
have expanded the total number of recorded glitches to 482
at the time of writing,
with up to 42 in an individual object (PSR J0537$-$6910).
Table \ref{tab:cor1} summarizes the size-waiting-time correlations
observed in the seven objects with $N \geq 10$ impulsive glitches,
an arbitrary cut-off.
The size of a glitch is defined by
$s=\Delta\nu/\nu $,
where $\Delta\nu$ is the instantaneous jump in pulse frequency $\nu$.
The forward (backward) waiting time from any given glitch 
to the next (previous) glitch
is denoted by $\Delta t_+$ ($\Delta t_-$).
For each object,
the table displays the Pearson coefficients 
\begin{equation}
 r_\pm 
 =
 \frac{\langle s \Delta t_\pm \rangle 
  - \langle s \rangle \langle \Delta t_\pm \rangle}
 {(\langle s^2 \rangle - \langle s \rangle^2)^{1/2}
  (\langle \Delta t_\pm^2 \rangle - \langle \Delta t_\pm \rangle^2)^{1/2}}
\label{eq:cor1}
\end{equation}
for the forward
($s$-$\Delta t_+$)
and backward ($s$-$\Delta t_-$) correlations
(where angular brackets denote an average),
the standard errors
\begin{equation}
 \sigma_{r_\pm} = 
 \left(
  \frac{1-r_\pm^2}{N-3}
 \right)^{1/2}
\label{eq:cor2}
\end{equation}
for the two correlations,
\footnote{
The factor $(N-3)^{-1/2}$ in (\ref{eq:cor2}) replaces the usual factor
$(N-2)^{-1/2}$, because $N$ glitches yield $N-1$ size-waiting-time pairs.
Likewise the PDF of $r_\pm / \sigma_{r_\pm}$ is a Student's t-distribution
with $N-3$ degrees of freedom, cf.\ $N-2$ usually.
\label{foot:cor2}
}
and the epoch $T_1$ of the first glitch in each sample.
For PSR J0534$+$2200 (Crab) and PSR J0835$-$4510 (Vela),
the correlations are computed for the full historical data set and
for subsets starting at Modified Julian Date (MJD) 46000,
when nearly continuous, single telescope monitoring began
using modern receivers and backends
\citep{lyn15}.
\footnote{
We include in the sample the latest glitch discovered in PSR J0534$+$2200,
which occurred at MJD 58237,
with $s=4.1\times 10^{-9}$
\citep{sha18b}.
}
For PSR J0537$-$6910,
the correlations are computed for the set of events that appear 
in at least two out of the three latest analyses
\citep{mid06,fer18,ant18}.
Several of the tabulated objects are likely to have experienced 
unpublished glitches in recent times; 
for example, PSR J0631$+$1036 glitched 15 times
from MJD 50186 to MJD 55702, yet no glitches have been published since then.
The data are plotted on a log-log scale in the form 
$s$ versus $\Delta t_+$ and $\Delta t_-$ 
in Figs \ref{fig:cor1}(a) and \ref{fig:cor1}(b)
respectively for the seven largest samples.

\begin{table}
\begin{center}
\begin{tabular}{lrcrcrcrcrc}
\hline
PSR J & $N$ & $T_1$ (MJD) & 
 $r_+$ & $\sigma_{r_+}$ & $r_-$ & $\sigma_{r_-}$ & 
 $\rho_+$ & $\sigma_{\rho_+}$ & $\rho_-$ & $\sigma_{\rho_-}$ \\
\hline
0534$+$2200 & 27 & 40493 &
 $-0.075$ & 0.204 & 0.328 & 0.193 & 0.024 & 0.204 & 0.464 & 0.181 \\
            & 23 & 46664 &
 $-0.113$ & 0.222 & 0.689 & 0.162 & $-0.095$ & 0.223 & 0.506 & 0.193 \\
0537$-$6910 & 42 & 51285 &
 0.927 & 0.060 & 0.159 & 0.158 & 0.931 & 0.058 & 0.164 & 0.158 \\
0631$+$1036 & 15 & 50186 &
 0.701 & 0.206 & $-0.091$ & 0.287 & 0.156 & 0.285 & $-0.145$ & 0.286 \\
0835$-$4510 & 21 & 40280 &
 0.407 & 0.215 & 0.603 & 0.188 & 0.358 & 0.220 & 0.398 & 0.216 \\
            & 14 & 46257 &
 0.607 & 0.240 & 0.661 & 0.226 & 0.484 & 0.264 & 0.533 & 0.255 \\
1341$-$6220 & 23 & 47989 &
 0.293 & 0.214 & $-0.082$ & 0.223 & 0.578 & 0.182 & $-0.145$ & 0.221 \\
1740$-$3015 & 35 & 47003 &
 0.298 & 0.169 & $-0.065$ & 0.176 & 0.264 & 0.171 & $-0.180$ & 0.174 \\
1801$-$2304 & 13 & 46907 &
 0.764 & 0.204 & $-0.024$ & 0.316 & 0.804 & 0.188 & $-0.042$ & 0.316 \\
\hline
\end{tabular}
\end{center}
\caption{
Size-waiting-time correlations and standard errors for 
actively glitching pulsars with $N \geq 10$.
}
\label{tab:cor1}
\end{table}

The random variables $r_\pm/\sigma_{r_\pm}$ follow a Student's t-distribution
with $N-3$ degrees of freedom (see footnote \ref{foot:cor2})
in the limit $N\rightarrow\infty$,
if the null hypothesis (zero correlation) is true,
and the underlying variables are drawn from a bivariate normal distribution.
The asymptotic result holds approximately for moderate $N$,
even if the underlying variables are not normally distributed.
Thus, as a first pass, we can test the null hypothesis,
that $s$ and $\Delta t_\pm$ are uncorrelated,
by computing the corresponding p-value
for the pulsars in Table \ref{tab:cor1},
i.e.\ the probability that the measured $|r_\pm|$ or greater arises by chance,
when the null hypothesis is true.
At 99.7 per cent confidence (three sigma),
using the full historical data set,
only one pulsar exhibits a significant $s$-$\Delta t_+$ correlation, 
namely PSR J0537$-$6910,
and zero pulsars exhibit a significant $s$-$\Delta t_-$ correlation.
At 95.4 per cent confidence (two sigma),
using the full historical data set,
PSR J0631$+$1036 and PSR J1801$-$2304 also exhibit significant 
$s$-$\Delta t_+$ correlations,
and PSR J0835$-$4510 exhibits a significant $s$-$\Delta t_-$ correlation.
\footnote{
Using the truncated data set, with $T_1 > {\rm MJD}\,46000$,
the following correlations are found in the Crab and Vela:
PSR J0534$+$2200 ($s$-$\Delta t_-$, three sigma),
PSR J0835$-$4510 ($s$-$\Delta t_+$ and $s$-$\Delta t_-$, two sigma).
}
Overall, the null hypothesis cannot be excluded for the majority of
the objects with $N\geq 10$,
even though it is tempting to see correlations other than those above
when inspecting Fig.\ \ref{fig:cor1} visually.

It may be argued that the Pearson correlation is not optimal
for glitch studies,
because (i) $s$ spans up to
four decades in individual objects (e.g.\ PSR J0534$+$2200),
biasing the covariance 
$\langle s \Delta t_\pm \rangle$
unduly towards events with the highest $s$;
and (ii) a nonlinear relation may exist between $s$ and $\Delta t_\pm$,
whereas the Pearson correlation tests for a linear relation.
For safety, therefore, we also calculate the Spearman rank correlation,
$\rho_\pm$,
which tests for a monotonic relation, whether it is linear or not,
and is less sensitive to outliers in the data.
The results including standard errors $\sigma_{\rho_\pm}$ are quoted 
in the last four columns of Table \ref{tab:cor1}.
It is clear by inspection that $r_\pm$ is broadly consistent with $\rho_\pm$
within the standard errors,
except possibly for the forward correlation in PSR J0631$+$1036,
where we find 
$| r_+ - \rho_+ | = 1.1(\sigma_{r_+} + \sigma_{\rho_+})$.
However,
upon rechecking the p-values, we discover that the forward correlation
for PSR J1801$-$2304 strengthens from two to three sigma;
the forward correlation 
for PSR J0631$+$1036 and the backward correlation for PSR J0835$-$4510
are no longer significant at 95.4 per cent confidence;
and new correlations arguably emerge for 
PSR J1341$-$6220 (forward; p-value $4.8\times 10^{-3}$)
\citep{yua10}
and 
PSR J0534$+$2200 (backward; p-value $1.7\times 10^{-2}$).
Therefore, in what follows, we adopt a conservative approach
and only deem correlations to be significant,
if they occur at the three-sigma level in the full historical data set,
i.e.\
$s$-$\Delta t_+$ for PSR J0537$-$6910 and PSR J1801$-$2304.
Detailed Monte Carlo simulations with realistic underlying PDFs
are deferred to a future paper, when more data become available
and warrant a more detailed study.
\footnote{
We also check and confirm that the results in Table \ref{tab:cor1}
are qualitatively unchanged, 
if we correlate $\Delta t_\pm$ against
${\rm log}_{10}s$ instead of $s$.
}

Quasiperiodic glitch activity is not accompanied always by
a strong $s$-$\Delta t_+$ correlation.
PSR J0537$-$6910 does glitch quasiperiodically,
but so does PSR J0835$-$4510, 
whose $s$-$\Delta t_+$ correlation is weak,
with Pearson and Spearman p-values $> 7.5\times 10^{-2}$
in the full data set
and $> 2.8\times 10^{-2}$ in the truncated data set.
This is interesting physically.
Quasiperiodic glitches are thought to occur,
when the stress reservoir empties almost completely at each event,
whereupon a strong $s$-$\Delta t_+$ correlation is expected.
Yet a strong $s$-$\Delta t_-$ correlation is also expected
under these circumstances,
and no pulsar in Table \ref{tab:cor1} exhibits it
at the three-sigma level in the full data set.

It is sometimes argued that the glitch sizes in PSR J0835$-$4510
are bimodally distributed
\citep{kon14,ash17}.
One can demonstrate,
using kernel density estimator techniques,
that the evidence for bimodality in glitch size PDFs is marginal 
for all the objects in Table \ref{tab:cor1}
\citep{how18}.
Nonetheless, for the sake of completeness,
we repeat the analysis
\footnote{
We thank G. Ashton for bringing this test and its results
to our attention.
}
for PSR J0835$-$4510 after excluding the three smallest glitches,
with $s \leq 10^{-7}$,
defined consistently as ``microglitches'' by \citet{pal16}.
We find
$r_+ = 0.421$, $r_- = 0.500$, 
$\rho_+ = 0.291$, and $\rho_- = 0.264$.
The latter coefficients are consistent with their
counterparts in Table \ref{tab:cor1};
the shifts lie well within the standard errors.
\footnote{
In a similar vein,
\citet{ant18} examined a subsample excluding the smallest events
in PSR J0537$-$6910 and found no significant differences in the
$s$-$\Delta t_\pm$ correlation coefficients;
see \S{4.1} and \S{4.2} in the latter reference.
}
At this juncture, the data yield no conclusive evidence for or against 
(i) the hypothesis of
two independent glitch mechanisms in PSR J0835$-$4510,
or (ii) a link between the $s$-$\Delta t_+$ correlation and quasiperiodicity.
We look forward to these matters being clarified in the future, 
when more data become available.

\section{State-dependent Poisson process
 \label{sec:cor3}}
We now show that the results in Table \ref{tab:cor1}
and Fig.\ \ref{fig:cor1} are consistent with an idealized yet general model 
of glitch activity as a state-dependent Poisson process
\citep{ful17}.
The model is not specific to a particular trigger mechanism.
It describes the system in the mean-field approximation
in terms of a global, random variable, $x(t)$,
which measures the spatially averaged differential rotation
(vortex avalanche picture)
or elastic stress
(starquake picture) throughout the star
as a function of time $t$.
As the star spins down, $x$ increases gradually,
at a rate proportional to the electromagnetic torque $N_{\rm em}$.
When a glitch occurs,
$x$ drops discontinuously by a random percentage.
Thus $x$ fluctuates around a mean value $\langle x \rangle$
over the long term.
Glitch triggering is postulated to be a Poisson process,
whose rate function $\lambda(x)$
(the number of trigger events per unit time)
increases monotonically with $x$.
As $x$ changes with time, so does $\lambda(x)$.

\begin{figure}
\begin{center}
\includegraphics[width=13cm,angle=0]{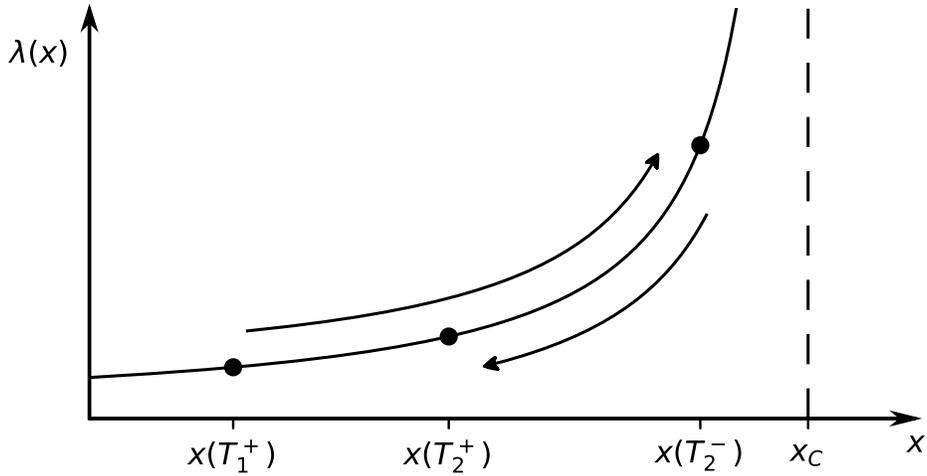}
\end{center}
\caption{
Schematic sketch of the
Poisson rate function, $\lambda(x)$, versus stress, $x$,
showing the divergence at the critical stress, $x_{\rm cr}$.
The epochs $T_n^-$ and $T_n^+$ occur
immediately before and after the $n$-th glitch respectively.
The stress increases gradually from $x(T_1^+)$ to $x(T_2^-)$
due to electromagnetic braking (rightward pointing arrow)
then drops discontinuously to $x(T_2^+)$ following a glitch at $t=T_2$ 
(leftward pointing arrow).
}
\label{fig:cor2}
\end{figure}

Consider a rate function $\lambda(x)$ 
of the form sketched in Fig.\ \ref{fig:cor2},
which diverges in the limit $x\rightarrow x_{\rm cr}$,
where $x_{\rm cr}$ is the critical stress.
The divergence is compatible with traditional glitch mechanisms
\citep{has15}.
In the vortex avalanche picture,
$x_{\rm cr}$ is the critical crust-superfluid angular velocity lag,
above which the Magnus force exceeds the pinning force throughout the star,
and every vortex unpins
\citep{lin91,ful17}.
\footnote{
Beyond the mean-field approximation, in a realistic star,
the threshold is exceeded earlier in some subregions,
so $x_{\rm cr}$ is a conservative upper limit.
}
In the starquake picture, 
$x_{\rm cr}$ is the critical elastic stress,
above which the crustal lattice fails catastrophically
\citep{mid06,hor09,chu10b,akb18}.
In the fluid instability picture,
$x_{\rm cr}$ is the critical relative velocity between superfluid components,
above which two-stream or Kelvin-wave instabilities are excited
\citep{and03,mas05,per06a,and07,gla09}.
Note that the rate in Fig.\ \ref{fig:cor2}
is small but nonzero in the limit $x \rightarrow 0$,
e.g.\ due to thermal activation
\citep{lin91}.

It is important to recognize that sizes and waiting times are likely to be
uncorrelated in the avalanche microphysics
underlying the vortex avalanche and starquake mechanisms.
This is a well-known property of any self-organized critical system
driven at a constant rate
\citep{jen98}.
For example, recent quantum mechanical, Gross-Pitaevskii simulations
of vortex avalanches in a pinned, decelerating Bose-Einstein condensate
show that the size of an avalanche is independent of the
crust-superfluid angular velocity lag $x$ immediately before the avalanche
\citep{war11,war13,mel15},
except that the avalanche size cannot exceed $x$, of course.
One can approach $x_{\rm cr}$ closely yet trigger a tiny avalanche;
counterintuitively, there is no tendency to trigger larger avalanches 
closer to the unpinning threshold.

Despite the absence of $s$-$\Delta t_\pm$ correlations at the microscopic level, 
such correlations do emerge, when uncorrelated 
avalanches are combined with global spin down.
To see this, consider rapid spin down firstly.
The system climbs rapidly up the $\lambda(x)$ curve in Fig.\ \ref{fig:cor2}
and almost reaches $x_{\rm cr}$, before a glitch occurs.
If $s$ is relatively large,
so is $|\Delta x|$,
the absolute value of the stress released by the glitch.
Hence $x$ faces a relatively long climb $\propto |\Delta x| / N_{\rm em}$
back to $x_{\rm cr}$ before the next glitch.
On the other hand, if $s$ is relatively small,
so is $|\Delta x|$,
and the delay $\propto |\Delta x | / N_{\rm em}$ until the next glitch
is relatively short.
This translates into a strong correlation between $s$ and $\Delta t_+$.
Note that the correlation emerges,
even though there is zero correlation between $|\Delta x|$
and the value of $x\approx x_{\rm cr}$ just before the glitch.
Also note that there is no significant $s$-$\Delta t_-$ correlation;
the time taken by $x$ to climb from its post-glitch starting point
up to $\approx x_{\rm cr}$ due to spin down has nothing to do with 
$| \Delta x |$ (and hence $s$) at the next glitch.

Next consider slow spin down.
Now the system does not reach $x \approx x_{\rm cr}$ before every glitch;
the avalanche is triggered at some intermediate value 
$x\approx \langle x \rangle$, 
with
$0 < \langle x \rangle < x_{\rm cr}$ 
and $\langle x \rangle \rightarrow 0$ as the spin-down rate decreases.
Hence the $s$-$\Delta t_+$ correlation in the previous paragraph
almost vanishes.
However, a weak $s$-$\Delta t_-$ correlation emerges instead.
The physics of the avalanche process is such that the stress variable $x$
cannot be negative, either for vortex avalanches or starquakes.
If the waiting time before a glitch is relatively short,
then $x$ is relatively small just before the glitch,
and so is the size of the avalanche, $| \Delta x | \leq x$;
i.e.\ $| \Delta x |$ is ``capped'', so that $x$ remains positive.
Conversely, if the waiting time is relatively long,
$| \Delta x |$ and hence $s$ can be larger while keeping $x$ positive always.
This translates into a weak correlation between $s$ and $\Delta t_-$.

\begin{figure}
\begin{center}
\includegraphics[width=11cm,angle=0]{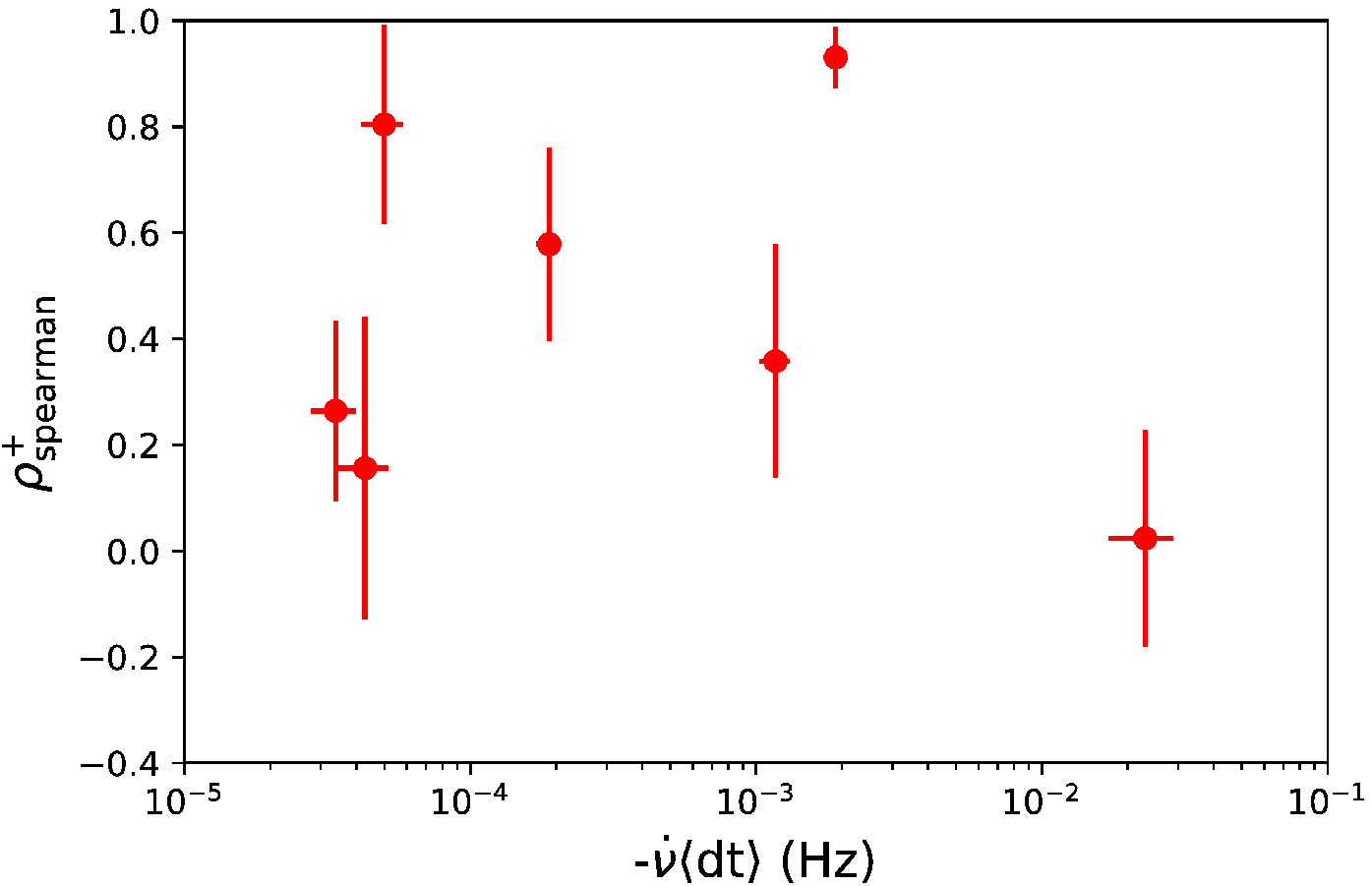}
\\
\vspace{1cm}
\includegraphics[width=11cm,angle=0]{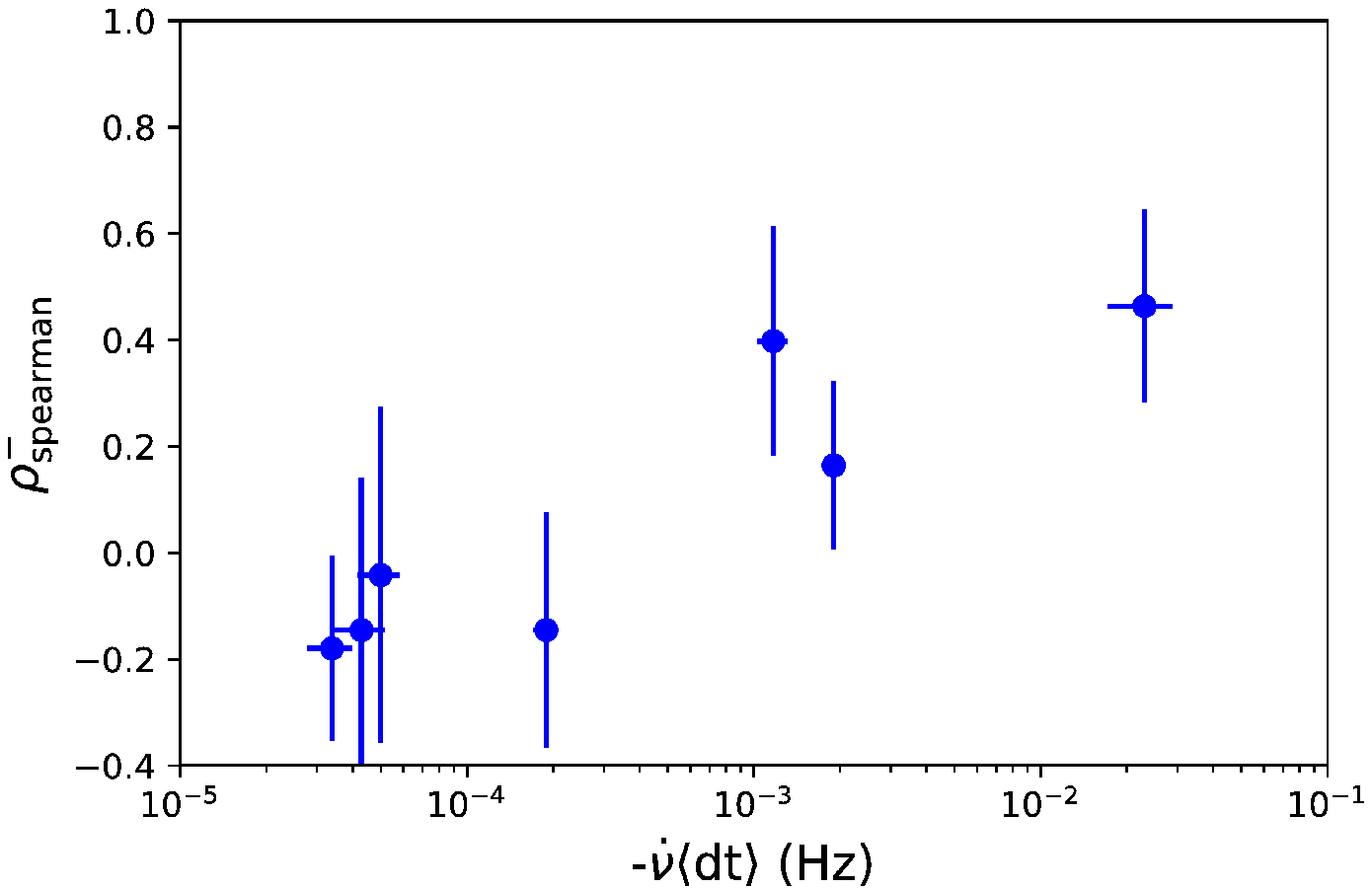}
\end{center}
\caption{
Spearman correlation coefficients $\rho_+$ 
(size versus forward waiting time; top panel, red symbols)
and $\rho_-$ 
(size versus backward waiting time; bottom panel, blue symbols)
as functions of the spin-down rate $-\dot{\nu}$ multiplied by 
the mean waiting time $\langle \Delta t \rangle$
(product in Hz)
for the pulsars in Table \ref{tab:cor1}.
Horizontal and vertical error bars are given by the
standard errors on $\langle \Delta t \rangle$ and $\rho_\pm$ respectively.
}
\label{fig:cor3}
\end{figure}

To test these ideas, we investigate how $r_\pm$ scales with
spin-down rate for the objects in Table \ref{tab:cor1}.
Immediately the question arises:
what measure of spin-down rate is it best to use?
The obvious candidate is $\dot{\nu}$, of course,
but $\dot{\nu}$ is clearly not the whole story;
if glitches occur frequently,
$\langle x \rangle$ can be much smaller than $x_{\rm cr}$,
even if $\dot{\nu}$ is large.
Another possibility is
$\dot{\nu} \langle \Delta t \rangle / x_{\rm cr}$,
which equals the mean stress accumulated between glitches
normalized by the critical stress.
The latter quantity has the advantages of being dimensionless
and equalling the reciprocal of one of the control parameters 
in the quantitative theory presented
in \S\ref{sec:cor4} 
[up to a factor of order unity;
 see \S\ref{sec:cor4c} and equation (\ref{eq:cor6})].
It has the disadvantage that $x_{\rm cr}$ is not observable.
We therefore compromise and 
plot $\rho_+$ (red symbols) and $\rho_-$ (blue symbols)
versus the dimensional yet observable quantity 
$-\dot{\nu} \langle \Delta t \rangle$
in Fig.\ \ref{fig:cor3}
for the seven objects in Table \ref{tab:cor1}.
We discuss the implications of the compromise carefully in \S\ref{sec:cor4c}
from a theoretical perspective.
The vertical error bars are given by $\sigma_{\rho_\pm}$,
while the horizontal error bars are given by the standard error of the mean,
$(N-1)^{-1/2} \sigma_{\Delta t}$,
where $\sigma_{\Delta t}$ is the standard deviation of the measured
waiting times.
\footnote{
The uncertainties in individual $\Delta t$ measurements range from
days to weeks for the objects in Table \ref{tab:cor1},
e.g.\ PSR J0537$-$6910 ($\leq 10\,{\rm days}$),
PSR J0631$+$1036 ($\leq 8\,{\rm days}$),
and PSR J1740$-$4015 ($\leq 24\,{\rm days}$ but mostly less than five days);
see \citet{mel08} for a detailed discussion
(specifically \S{3}, paragraph 4 in \S{5.1}, and Table 1 in the
latter reference).
However, for small samples with $N\leq 35$,
the dispersion from individual uncertainties is modest
compared to the standard error of the mean,
which typically exceeds one month for the objects in Table \ref{tab:cor1}.
}
The measurement uncertainty in $\dot{\nu}$ is negligible.
Note that $\dot{\nu}$ is the long-term, average, spin-down rate
after correcting for glitches and timing noise,
as quoted in the Australia Telescope National Facility Pulsar Catalogue
\citep{man05}.

One result stands out from Fig.\ \ref{fig:cor3}:
the strongest $s$-$\Delta t_+$ correlation found in the sample
is associated with PSR J0537$-$6910,
which has the second-highest $-\dot{\nu}\langle \Delta t \rangle$ 
among the plotted objects.
This is consistent with the behavior predicted above
for a state-dependent Poisson process with $\lambda(x)$ qualitatively
of the form sketched in Fig.\ \ref{fig:cor2}.
Moreover, PSR J0537$-$6910 exhibits no statistically significant
$s$-$\Delta t_-$ correlation,
which also matches the predicted behavior of a state-dependent
Poisson process.
Beyond that, the picture is cloudy.
PSR J0534$+$2200 has the largest $-\dot{\nu}\langle \Delta t \rangle$ 
in the sample by $\approx 1\,{\rm dex}$,
yet it exhibits no significant $s$-$\Delta t_+$ correlation
\citep{won01,esp14,sha18}
and, if anything, exhibits a two-sigma $s$-$\Delta t_-$ correlation
according to the Spearman test.
PSR J1801$-$2304 does exhibit a three-sigma $s$-$\Delta t_+$ correlation,
yet it has the third-lowest $-\dot{\nu} \langle \Delta t \rangle$
in the sample.
It is hard to know what to make of these results without dividing
$-\dot{\nu} \langle \Delta t \rangle$ by $x_{\rm cr}$,
but $x_{\rm cr}$ is unknown and varies in general from pulsar to pulsar.
We therefore postpone discussion of the less statistically significant
features of Fig.\ \ref{fig:cor3}, until more data become available,
and a better understanding of $x_{\rm cr}$ in specific objects develops.

\section{Quantitative analysis
 \label{sec:cor4}}
To prepare for the arrival of more data,
we predict the size-waiting-time correlation theoretically in this section. 
The calculation follows directly from the theory
of a state-dependent Poisson process developed by \citet{ful17}
for glitches triggered by superfluid vortex avalanches.
It applies equally to starquakes for the reasons expressed in \S\ref{sec:cor3}.

\subsection{Equations of motion
 \label{sec:cor4a}}
In general,
the stress variable $x(t)$ obeys a stochastic equation of motion
of the form \citep{ful17}
\begin{equation}
 x(t) = x(0) + t - \sum_{i=1}^{N(t)} \Delta x^{(i)}~,
\label{eq:cor3}
\end{equation}
where $x(0)$ is an astrophysically irrelevant initial stress,
the second term on the right-hand side describes the secular increase
in $x$, as the star spins down,
$N(t)$ is the number of glitches having occurred up to time $t$,
and $\Delta x^{(i)}$ is the absolute value of the step decrease in $x$
due to the $i$-th glitch.
Equation (\ref{eq:cor3}) is written in dimensionless form,
with $x$ and $\Delta x^{(i)}$ expressed in units of $x_{\rm cr}$,
and $t$ expressed in units of $x_{\rm cr} I_{\rm c}/N_{\rm em}$,
where $I_{\rm c}$ is the moment of inertia of the stellar crust;
see \S{3.4} of \citet{ful17} for details.
\footnote{
A different normalization for $t$ is needed in the starquake picture, 
where $x$ has the units of elastic stress rather than angular velocity.
}

Random processes like (\ref{eq:cor3}) are called doubly stochastic
\citep{cox55},
because both $N(t)$ and $\Delta x^{(i)}$ are random variables.
In between two glitches, 
in the interval $t_{\rm g} \leq t' \leq t_{\rm g}+\Delta t$,
the dimensionless stress evolves deterministically according to
$x(t') = x(t_{\rm g}^+) + t'-t_{\rm g}$ (i.e.\ spin down),
where $x(t_{\rm g}^+)$ denotes the stress immediately after the first glitch.
The PDF of the waiting time $\Delta t$ obeys the classic formula
for a time-dependent Poisson process, viz.\
\begin{equation}
 p[ \Delta t | x(t_{\rm g}^+) ]
 =
 \lambda[ x(t_{\rm g}^+) + \Delta t ]
 \exp\left\{
  -\int_{t_{\rm g}}^{t_{\rm g}+\Delta t}
  dt' \, \lambda [x(t')]
 \right\}~.
\label{eq:cor4}
\end{equation}
Following \citet{ful17}, we work with the rate function
\begin{equation}
 \lambda(x)
 =
 \frac{\alpha}{1-x}~,
\label{eq:cor5}
\end{equation}
where
\begin{equation}
 \alpha
 =
 \frac{I_{\rm c} x_{\rm cr} \lambda_0}{N_{\rm em}}
\label{eq:cor6}
\end{equation}
is a dimensionless control parameter proportional to the 
microscopic avalanche trigger (e.g.\ vortex unpinning) rate $2\lambda_0$
at the reference stress $x=x_{\rm cr}/2$.
\footnote{
Equivalently $\lambda_0$ is the trigger rate at zero stress,
but it is safer to think of it as a characteristic rate at $x=x_{\rm cr}/2$,
just in case the physics at $x=0$ (e.g.\ thermal activation) 
is radically different to the physics at $x \sim x_{\rm cr}$.
}
Equation (\ref{eq:cor6}) embodies the properties discussed in \S\ref{sec:cor3}
and sketched in Fig.\ \ref{fig:cor2}.
Its specific, hyperbolic functional form is arbitrary;
the results do not depend sensitively on it,
e.g.\ $\lambda(x) = 2\alpha \tan (\pi x /2)$ 
works just as well
\citep{ful17}.
The PDF of the jump sizes $\Delta x^{(i)} > 0$ is given by
the conditional jump probability
\begin{equation}
 \eta(x|y)
 \propto
 (y-x)^{-1.5} H(y-x-\beta y)~,
\label{eq:cor7} 
\end{equation}
where $\eta(x|y)\, dx$ equals the probability of jumping from $y$ to 
a stress value in the interval $(x,x+dx)$,
with $y-\Delta x^{(i)} = x$ at the $i$-th glitch.
Every glitch reduces the stress,
the Heaviside function $H(\dots)$ in (\ref{eq:cor7}) ensures that
no glitch makes $x$ negative,
and the minimum stress release is $\beta y$ ($0<\beta < 1$)
(required for normalization).
The power-law form and exponent of (\ref{eq:cor7})
are chosen to be consistent with the avalanche size PDFs seen universally
in self-organized critical systems like sandpiles, earthquakes, and solar flares
\citep{jen98,sor04,asc16}
and specifically in Gross-Pitaevskii simulations of superfluid
vortex avalanches in the neutron star context
\citep{war11,mel15}.
Monte Carlo simulations confirm that,
as with $\lambda(x)$, the output of the model does not depend sensitively
on the specific functional form of $\eta(x|y)$
\citep{ful17}.
There is no way at present of measuring $\eta(x|y)$ observationally
or deriving it theoretically from first principles.
A new generation of Gross-Pitaevskii simulations containing many more vortices
than have been analysed to date would be required,
a challenging computational task.

\subsection{Critical spin-down rate
 \label{sec:cor4b}}
The behavior of the model (\ref{eq:cor3})--(\ref{eq:cor7})
was studied thoroughly as a function of the control parameters
$\alpha$ and $\beta$ by \citet{ful17}
using Monte Carlo simulations and analytic theory.
The behavior divides into two distinct regimes:
large $\alpha \gtrsim \alpha_{\rm c}(\beta)$ (slow spin down)
and small $\alpha \lesssim \alpha_{\rm c}(\beta)$ (fast spin down),
with
\begin{equation}
 \alpha_{\rm c} \approx \beta^{-1/2}~.
\label{eq:cor8} 
\end{equation}
In the large-$\alpha$ regime,
the simulations produce power-law and exponential PDFs
for $s$ and $\Delta t$ respectively
[see Figs 6 and 8 respectively in \citet{ful17}],
consistent with observations of many pulsars
\citep{mel08,esp11,ash17,how18}.
In the small-$\alpha$ regime,
$p(s)$ and $p(\Delta t)$ have nearly the same functional form,
i.e.\ $p(s)\approx p(\Delta t)$ in terms of dimensionless variables,
which is consistent with observations of quasiperiodic objects,
except that the functional form is a power law instead of a Gaussian
for the specific jump distribution (\ref{eq:cor7}).

\subsection{$s$-$\Delta t_\pm$ correlations
 \label{sec:cor4c}}
\begin{figure}
\begin{center}
\includegraphics[width=13cm,angle=0]{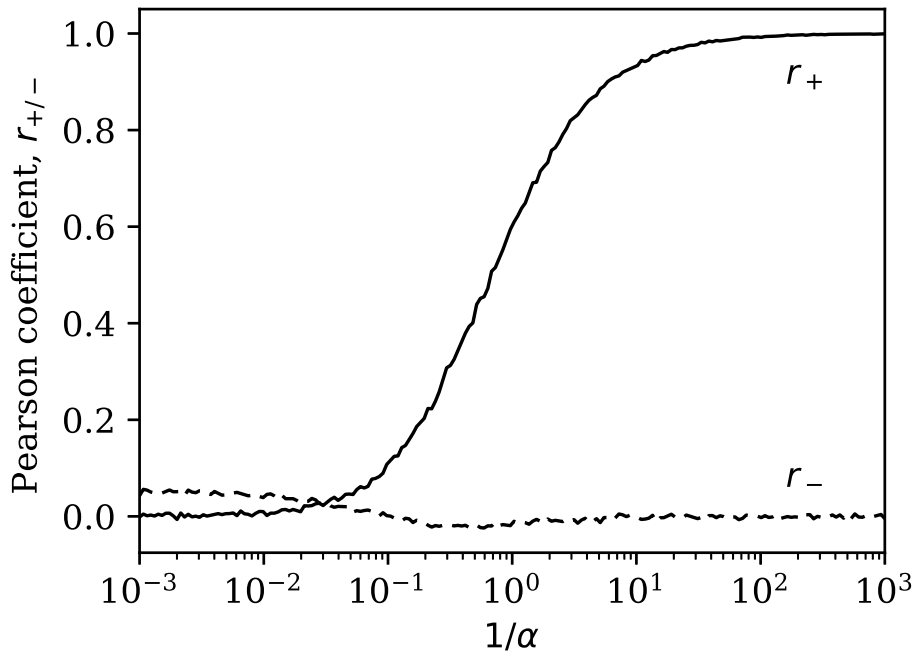}
\end{center}
\caption{
Theoretical Pearson correlation coefficients $r_+$ 
(size versus forward waiting time; solid curve)
and $r_-$ 
(size versus backward waiting time; dashed curve)
as functions of the control parameter $\alpha^{-1}$ 
[equation (\ref{eq:cor6})],
generated by Monte Carlo simulations of the model
(\ref{eq:cor3})--(\ref{eq:cor7}) with $\beta=10^{-2}$
[see \citet{ful17}].
Simulation parameters:
200 logarithmically spaced $\alpha$ values,
$10^5$ glitches per $\alpha$ value.
The Spearman coefficients are not graphed
because they are harder to compare against the analytic theory
in Appendix \ref{sec:corappa}.
}
\label{fig:cor4}
\end{figure}

Just as $p(s)$ and $p(\Delta t)$ change character at
$\alpha \approx \alpha_{\rm c}(\beta)$,
so do the $s$-$\Delta t_\pm$ correlations.
Fig.\ \ref{fig:cor4} displays $r_\pm$ versus $\alpha$
for the jump distribution (\ref{eq:cor7}).
The plot spans the full range from small to large $\alpha$,
with $\beta = 10^{-2}$ and hence $\alpha_{\rm c} \approx 10$
in this example.
The behavior exactly matches what is predicted by the 
qualitative discussion in \S\ref{sec:cor3}.
A strong forward correlation emerges,
when the spin-down rate is fast,
because the stress approaches $x\approx x_{\rm cr}$ before every glitch.
A weak backward correlation emerges,
when the spin-down rate is slow,
because the size of a glitch is capped to ensure $x\geq 0$.
The transition occurs at $\alpha \approx \alpha_{\rm c}$ 
in Fig.\ \ref{fig:cor4}.

Do the measured values of $r_\pm$ agree with the theoretical prediction
in Fig.\ \ref{fig:cor4}?
To answer this question, we need to know $\alpha$ 
for the objects in Table \ref{tab:cor1}.
Unfortunately,
equation (\ref{eq:cor6}) expresses $\alpha$ in terms of the quantities
$I_{\rm c}$, $x_{\rm cr}$, $\lambda_0$, and $N_{\rm em}$,
none of which can be measured directly.
We can write $2\pi \dot{\nu} \approx N_{\rm em}/I_{\rm c}$
to a good approximation, 
because the electromagnetic braking torque
dominates the superfluid back-reaction torque on the crust
\citep{esp11}.
However $x_{\rm cr}$ and $\lambda_0$ cannot be related easily
to non-glitch observables.
We therefore turn to the theoretical analysis in Appendix \ref{sec:corappa}
for inspiration.
Although it applies to the special case where $\eta(x|y)$ is separable,
nevertheless it turns out to offer useful clues.
From (\ref{eq:corappa13}), we find that $\alpha$ can be related
to the observable mean waiting time,
$\langle \Delta t \rangle$,
via 
$\alpha \approx 
 x_{\rm cr} / (2\pi \langle \Delta t \rangle \dot{\nu})$
up to a proportionality factor of order unity,
where we now restore the dimensions to $\langle \Delta t \rangle$.
Clearly $\lambda_0$ drops out of the expression, leaving $x_{\rm cr}$.
Suppose we then make the assumption,
that $x_{\rm cr}$ does not vary much from one pulsar to the next,
because it is set by the balance of the Magnus and pinning forces
(vortex avalanche picture) or crustal breaking strain (starquake picture),
which are nuclear in origin and independent of the rotational state
($\nu$, $\dot{\nu}$).
Then $\alpha$ is inversely proportional to the observable product
$\dot{\nu} \langle \Delta t \rangle$,
and it is possible to use this product to compare $r_\pm$ across
different pulsars.
This motivates the choice of normalization of the abscissae 
in Fig.\ \ref{fig:cor3}, as foreshadowed in \S\ref{sec:cor3}.

The crude first success of Fig.\ \ref{fig:cor3} --- 
that the object with the highest $r_+$ also happens to have
the second-highest value of 
$-\dot{\nu} \langle \Delta t \rangle$,
in line with the theory ---
is an encouraging sign that a stick-slip process
described by (\ref{eq:cor3}) may be at work.
However it is nothing more than a first indication;
many more data are needed, before we can say anything definite.
Certainly the assumption in the previous paragraph,
that $x_{\rm cr}$ does not vary much from one pulsar to the next,
is unlikely to hold exactly.
Variation in $x_{\rm cr}$ between objects is one natural way to explain
why PSR J0534$+$2200 fails to exhibit a strong $s$-$\Delta t_+$ correlation,
despite having the highest
$-\dot{\nu} \langle \Delta t \rangle$ in Table \ref{tab:cor1};
$x_{\rm cr}$ may be larger than average in this pulsar.
Likewise,
PSR J1801$-$2304 does exhibit a strong $s$-$\Delta t_+$ correlation,
even though it has the third-lowest
$-\dot{\nu} \langle \Delta t \rangle$
in Table \ref{tab:cor1};
$x_{\rm cr}$ may be smaller than average in this pulsar.

The reader might wonder whether some other observables,
e.g.\ $\langle s \rangle$ or $\langle (\Delta t )^2 \rangle$,
depend on $\alpha$ and $x_{\rm cr}$ in a different combination,
allowing us to disentangle the values of $\alpha$ and $x_{\rm cr}$.
Unfortunately the prospects are dim.
One finds from (\ref{eq:corappa13}) and (\ref{eq:corappa25})
that the dimensionless ratio
$\langle (\Delta t )^2 \rangle / \langle \Delta t \rangle^2$
depends primarily on $\alpha$, but the dependence is weak,
and $\alpha$ is poorly constrained given the measurement uncertainties.
\footnote{
From the Appendix we have
${\rm var} (\Delta t ) / \langle \Delta t \rangle^2
 = (\alpha+\delta+1)/(\alpha+\delta+3)$,
independent of $x_{\rm cr}$.
The parameter $\delta \approx 3$ in the unmeasurable jump distribution
introduces another uncertainty.
}
Likewise long-term conservation of angular momentum
implies $\langle s \rangle / \langle \Delta t \rangle = 1$
upto a factor involving the crust and superfluid moments of inertia,
and again $x_{\rm cr}$ cannot be disentangled;
we have $\langle s \rangle \propto x_{\rm cr}$ and
$\langle \Delta t \rangle \propto x_{\rm cr}$,
and hence $x_{\rm cr}$ cancels out in the ratio.

\section{Targets
 \label{sec:cor5}}
We conclude by using the results in \S\ref{sec:cor3} and \S\ref{sec:cor4}
to predict what pulsars are likely
to display strong size-waiting-time correlations in the future,
when more data become available.

\begin{figure}
\begin{center}
\includegraphics[width=13cm,angle=0]{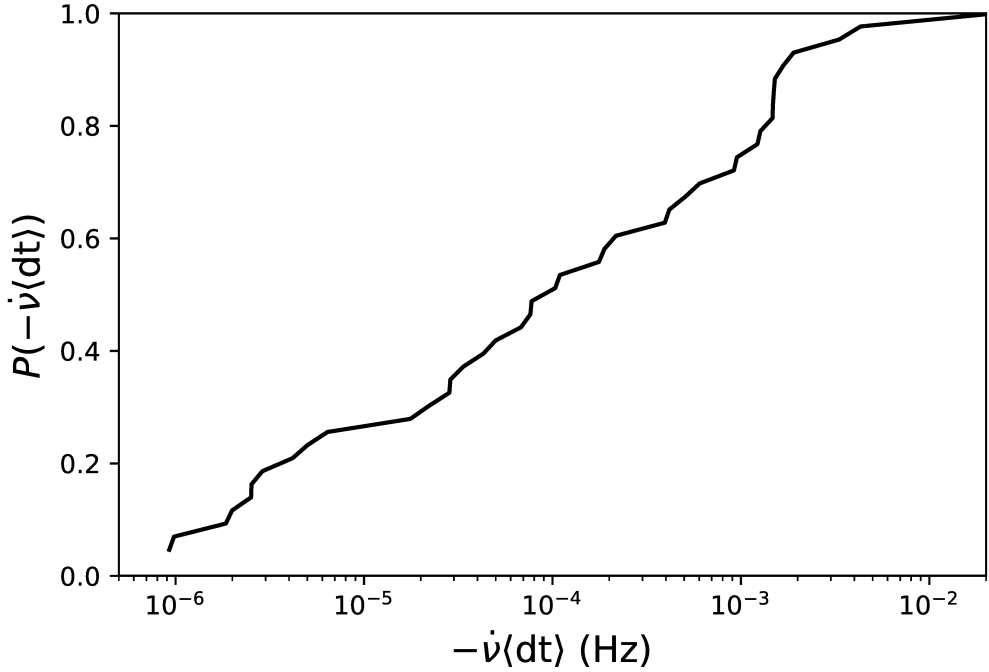}
\end{center}
\caption{
Cumulative distribution function of $-\dot{\nu} \langle \Delta t \rangle$
for glitching pulsars with $N\geq 4$.
}
\label{fig:cor5}
\end{figure}

In Fig.\ \ref{fig:cor5} we present the cumulative distribution function
of $-\dot{\nu} \langle \Delta t \rangle$
for all pulsars known to glitch at the time of writing with $N\geq 4$,
so that $\sigma_{r_\pm}$ is well defined.
In Table \ref{tab:cor2} we name the pulsars
with the five highest and five lowest
$-\dot{\nu} \langle \Delta t \rangle$ values.
From the results in \S\ref{sec:cor3} and \S\ref{sec:cor4},
we venture to make two predictions.
First,
if $r_+$ is measured to be high in a pulsar,
then that particular object is likely to lie towards the top end of the 
$-\dot{\nu} \langle \Delta t \rangle$ distribution,
depending on its $x_{\rm cr}$ value.
By and large, therefore,
the objects in the top half of Table \ref{tab:cor2} represent
good $r_+$ targets
(except for PSR J0534$+$2200; see Table \ref{tab:cor1}).
Second,
we predict that no glitching pulsar will exhibit a strong
$s$-$\Delta t_-$ correlation, either now or in the future.
Equation (\ref{eq:corappa26}) implies $r_- \leq 0.5$
for separable $\eta(x|y)$
and $r_- \lesssim 0.1$ for typical parameters.
Among the low $r_-$ measurements,
we predict that the highest will lie towards the bottom end of the
$-\dot{\nu} \langle \Delta t \rangle$ distribution,
again depending on $x_{\rm cr}$.
By and large, 
the objects in the bottom half of Table \ref{tab:cor2} represent
good $r_-$ targets.
For the sake of completeness, 
we quote $r_\pm$ in the last two columns of the table,
as computed from existing data.
However, we urge the reader not to draw any conclusions at this stage
about objects other than PSR J0534$+$2200 and PSR J0537$-$6910
in Table \ref{tab:cor2};
the samples are simply too small ($4\leq N \leq 7$)
to say anything with confidence.

We emphasize that the above predictions implicitly assume,
that $x_{\rm cr}$ does not vary much from one object to the next 
(see \S\ref{sec:cor4}),
so that $\alpha$ and $\dot{\nu} \langle \Delta t \rangle$ 
can be used interchangeably.
This seems unlikely, when one considers the nuclear physics of the crust,
and may well explain the existing misfits PSR J0534$+$2200 
(high $-\dot{\nu} \langle \Delta t \rangle$, low $r_+$)
and PSR J1801$-$2304
(low $-\dot{\nu} \langle \Delta t \rangle$, high $r_+$).
The predictions also assume, that $\beta$ and hence
$\alpha_{\rm c}\approx \beta^{-1/2}$
do not vary much between pulsars,
which is an open question physically.
Therefore the predictions should be seen as a first step towards
falsifiable tests of the correlation mechanism,
to be refined
as our understanding of $x_{\rm cr}$ and $\beta$ in specific objects improves.

\begin{table}
\begin{center}
\begin{tabular}{lrcrr}
\hline
PSR J & $N$ & $-\dot{\nu}\langle \Delta t \rangle$ (Hz) & $r_+$ & $r_-$ \\
\hline
0205$+$6449 & 6  & $3.32\times 10^{-3}$ & 0.947
 & $-0.447$ \\
0534$+$2200 & 27 & $2.23\times 10^{-2}$ & $-0.075$
 & 0.328 \\
0537$-$6910 & 42 & $1.90\times 10^{-3}$ & 0.927
 & 0.159 \\
1119$-$6127 & 4  & $4.31\times 10^{-3}$ & 0.900
 & 0.860 \\
2229$+$6114 & 6  & $1.68\times 10^{-3}$ & 0.874
 & $-0.305$ \\
\hline
0528$+$2200 & 4 & $9.81\times 10^{-7}$ & $-0.645$
 & 0.768 \\
1814$-$1744 & 7 & $2.89\times 10^{-6}$ & 0.042
 & 0.222 \\
1902$+$0615 & 6 & $1.99\times 10^{-6}$ & 0.490
 & $-0.314$ \\
1957$+$2831 & 4 & $4.18\times 10^{-6}$ & 0.667
 & 0.613 \\
2225$+$6535 & 5 & $5.01\times 10^{-6}$ & 0.998
 & $-0.325$ \\
\hline
\end{tabular}
\end{center}
\caption{
Glitching pulsars with the highest (top) and lowest (bottom) values of
$- \dot{\nu} \langle \Delta t \rangle$ and $N\geq 4$
proposed as targets for future correlation analyses,
when more data are gathered.
The current values of the Pearson coefficients $r_\pm$ are tabulated
for completeness, but most of the samples are too small
($4\leq N \leq 7$) to draw any statistically significant conclusions.
}
\label{tab:cor2}
\end{table}

As more data become available,
it will be possible in principle to turn around the above predictions
and use measured correlations (or their absence) to constrain $x_{\rm cr}$
and the stress-release physics in glitches.
Consider $r_+$ for example.
Every glitching pulsar that is measured to have $r_+ \ll 1$
also has $\alpha \gtrsim \alpha_{\rm c}$ 
and hence
$\beta^{1/2} x_{\rm cr} \gtrsim - 2\pi \dot{\nu} \langle \Delta t \rangle$
in the theory in \S\ref{sec:cor4},
upon relating $\alpha$ to $x_{\rm cr}$ as before and using (\ref{eq:cor8}).
To illustrate what is possible in the future,
we note that we obtain minimum values of $\beta^{1/2} x_{\rm cr} / (2\pi)$
between $2.3\times 10^{-2}\,{\rm Hz}$ and $3.4\times 10^{-5}\,{\rm Hz}$
for the objects in Table \ref{tab:cor1} with $r_+ \ll 1$
and between $2.3\times 10^{-2}\,{\rm Hz}$ and $9.8\times 10^{-7}\,{\rm Hz}$
for the objects in Table \ref{tab:cor2}.
These bounds are consistent with sensible values of
$x_{\rm cr}$ and $\beta$ in the vortex avalanche picture
\citep{lin91,war11}, e.g.
\begin{equation}
 \beta^{1/2} x_{\rm cr} / (2\pi)
 = 1 \times 10^{-3}
 \left( \frac{\beta}{10^{-2}} \right)^{1/2}
 \left( \frac{F_{\rm max}}{\rm keV\,fm^{-1}} \right)
 \left( \frac{\rho}{10^{13}\,{\rm g\,cm^{-3}}} \right)^{-1}
 \left( \frac{l}{10^2\,{\rm fm}} \right)^{-1}
 {\rm Hz}~,
\label{eq:cor9}
\end{equation}
where $F_{\rm max}$ is the maximum pinning force per site,
$\rho$ is the superfluid density,
and $l$ is the pinning site separation.
An analogous expression in the starquake picture can be deduced
from the models in \citet{mid06} and \citet{akb18}.
In the vortex avalanche picture especially,
a lot of complicated physics goes into $F_{\rm max}$,
including the form of the nuclear pinning potential,
vortex tension, single- versus multi-site breakaway,
and collective avalanche knock-on;
see \citet{has15} and references therein.
One therefore expects $F_{\rm max}$ to vary
from one pulsar to the next.

\section{Conclusion
 \label{sec:cor6}}
In this paper, we quantify systematically the size-waiting-time correlations
observed in pulsar glitches using the Pearson and Spearman coefficients.
We find that, at the three-sigma level,
no objects exhibit a significant $s$-$\Delta t_-$ correlation,
and only two, PSR J0537$-$6910 and PSR J1801$-$2304, 
exhibit significant $s$-$\Delta t_+$ correlations.
We show that these results can be understood theoretically
in terms of a state-dependent Poisson process,
whose rate diverges when the system stress approaches a critical threshold
$x_{\rm cr}$ in both the vortex avalanche and starquake pictures.
The state-dependent Poisson process predicts a strong $s$-$\Delta t_+$
correlation ($r_+ \approx 1$) for fast spin down,
i.e.\ for $-\dot{\nu} \langle \Delta t \rangle / x_{\rm cr}$
greater than a critical value related to the minimum avalanche size.
It also predicts a weak $s$-$\Delta t_-$ correlation ($r_- \approx 0$) 
for fast and slow spin down.
Applying the theory to the list of known, glitching pulsars with $N\geq 4$,
ranked by $-\dot{\nu} \langle \Delta t \rangle$,
we identify the objects that are likely to display strong $s$-$\Delta t_+$
correlations (and weak or nonexistent $s$-$\Delta t_-$ correlations),
as more data are collected.
The prediction relies to some extent on assuming that $x_{\rm cr}$ 
and the minimum avalanche size, which are unobservable,
do not vary much from one pulsar to the next.
If future data are in accord with this assumption, 
measurements of $r_\pm$ versus $-\dot{\nu}\langle\Delta t \rangle$
can be turned around
to constrain $x_{\rm cr}$ and hence
the nuclear pinning forces (vortex avalanche picture)
or crustal breaking strain (starquake picture) in individual pulsars.

The results in this paper extend the theoretical framework developed
by \citet{ful17} by focusing on size-waiting-time
correlations as a quantitative observational test of the model.
The new elements include:
(i) a systematic, multi-object analysis of the Pearson and Spearman 
coefficients derived from data in the Jodrell Bank and Australia Telescope 
glitch catalogues (\S\ref{sec:cor2});
(ii) intuitive explanations for the strong $s$-$\Delta t_+$
and weak $s$-$\Delta t_-$ correlations expected in a 
state-dependent Poisson process (\S\ref{sec:cor3});
(iii) closed form integral expressions for $r_\pm$ 
(\S\ref{sec:corappab});
(iv) a recipe for relating the correlation data to 
essential nuclear physics parameters, e.g.\ maximum pinning force
[\S\ref{sec:cor4c} and equation (\ref{eq:cor9})];
and (v) predictions for what specific pulsars are most likely
to exhibit emerging $s$-$\Delta t_\pm$ correlations,
as more observations are made.

We emphasize again in closing that the theoretical framework
is not specific to a particular version of the glitch microphysics.
The state-dependent Poisson process is a meta-model which encompasses
all the glitch mechanisms contemplated in the literature to date,
e.g.\ starquakes and superfluid vortex avalanches.
It rests on two assumptions of a general nature:
(i) the stress $x$ increases gradually between glitches and
relaxes discontinuously at a glitch;
and (ii) the trigger rate $\lambda(x)$ increases with $x$
and diverges at $x_{\rm cr}$.
If the meta-model is falsified in the future,
with the arrival of more data and a better understanding of $x_{\rm cr}$
in specific objects,
a fresh approach to the glitch problem will be required.

In order to take full advantage of the opportunity for falsification,
more glitches need to be found.
Improved data analysis techniques will play an important role
in this regard.
Recent innovations
include algorithms that harness the power of
distributed volunteer computing
\citep{cla17},
alternatives to least-squares fitting for nongaussian noise
\citep{wan17},
and Bayesian model selection
\citep{sha16}.

\acknowledgments
The authors thank Julian Carlin for assistance with the preparation
of Figs \ref{fig:cor2} and \ref{fig:cor4}.
This research was supported by the Australian Research Council
Centre of Excellence for Gravitational Wave Discovery (OzGrav),
grant number CE170100004.
GH is the recipient of an Australian Postgraduate Award.
This work was performed in part at the Aspen Center for Physics,
which is supported by National Science Foundation grant PHY--1607611.

\bibliographystyle{mn2e}
\bibliography{glitchstat}

\appendix
\section{Glitch master equation
 \label{sec:corappa}}
In this appendix, we summarize certain useful results from
an analytic theory developed by \citet{ful17}
to predict the long-term glitch statistics generated by
(\ref{eq:cor3})--(\ref{eq:cor7}).
The aims are to justify the theoretical relations between observables 
(e.g.\ $r_\pm$ and $\dot{\nu}\langle \Delta t \rangle$)
discussed in \S\ref{sec:cor3} onwards
and motivate the axis choices made in Fig.\ \ref{fig:cor3} onwards.

For $t\geq 1-x(0)$, the system (\ref{eq:cor3})--(\ref{eq:cor7})
exhibits stationary behavior:
$x(t)$ fluctuates about a constant mean,
$0 < \langle x \rangle < 1$,
governed by the balance between the second and third terms on the
right-hand side of (\ref{eq:cor3}).
The system is self-regulating,
because $\lambda(x)$, which determines $N(t)$,
increases monotonically with $x$;
as the stress rises, glitches occur more frequently and relax the system.
Under stationary conditions,
the PDF $p(x)$ of the stress variable $x$
satisfies the time-independent master equation
\citep{war13,ful17},
\begin{equation}
 0 =
 -\frac{d p(x)}{d x}
 - \lambda(x) p(x)
 + \int_x^1 dy \, p(y) \lambda(y) \eta(x|y)~.
\label{eq:corappa1}
\end{equation}
Equations (\ref{eq:corappa1}) and (\ref{eq:cor3}) describe
exactly the same dynamics 
and are expressed in terms of the same dimensionless variables.
The first two terms on the right-hand side of (\ref{eq:corappa1})
describe the probability lost from the interval $(x,x+dx)$
due to secular spin down and discontinuous jumps (glitches)
out of the interval respectively.
The third term describes the integrated probability gained
in the interval $(x,x+dx)$,
when glitches take the system from another state $y$ into $(x,x+dx)$.
Once $p(x)$ is known after solving (\ref{eq:corappa1}), 
it is possible to calculate the statistical distributions of
other system variables,
including observables like $s$ and $\Delta t_\pm$.

Equations (\ref{eq:corappa1}) and (\ref{eq:cor5})--(\ref{eq:cor7})
form a closed system, which can be solved by the methods developed
by \citet{ful17}. 
Monte Carlo simulations confirm that the solution is insensitive 
to the particular choices of $\lambda(x)$ and $\eta(x|y)$, 
as the latter reference demonstrates.
If $\eta(x|y)$ is separable, the theory can even be solved analytically.
In this appendix, we present the analytic solution for
\begin{equation}
 \eta(x|y)
 =
 (\delta + 1)
 x^\delta y^{-(\delta+1)}~.
\label{eq:corappa2}
\end{equation}
This choice is illustrative only; 
the vortex or starquake avalanche dynamics inside
a neutron star cannot be measured experimentally at present.
However it is consistent with the output of Gross-Pitaevskii simulations,
viz.\ equation (\ref{eq:cor7})
\citep{war11},
and correctly favors small avalanches over large ones for $\delta > 0$,
with $\delta \approx 3$ yielding event statistics broadly
in accord with those generated by (\ref{eq:cor7}).
It also leads to generic scalings between observables,
which are reproduced by other sensible choices of $\eta(x|y)$ too,
as confirmed by Monte Carlo simulations with nonseparable $\eta(x|y)$
performed by \citet{ful17}.

\subsection{Stress, size, and waiting-time PDFs
 \label{sec:corappaa}}
Solving (\ref{eq:corappa1}) and (\ref{eq:corappa2})
by separation of variables, as in Appendices C and D in \citet{ful17},
we find
\begin{equation}
 p(x)
 =
 C x^{\delta+1} (1-x)^\alpha~,
\label{eq:corappa3}
\end{equation}
with
\begin{equation}
 C = \frac{\Gamma(\alpha+\delta+3)}{\Gamma(\alpha+1) \Gamma(\delta+2)}~,
\label{eq:corappa4}
\end{equation}
where $\Gamma(\dots)$ symbolizes the gamma function.
Equation (\ref{eq:corappa3}) implies
$0 < \langle x \rangle = (\delta + 2)(\alpha + \delta + 3)^{-1} < 1$.
We can also calculate the PDFs of $x$ immediately before and after a glitch,
called $p_{\rm e}(x)$ and $p_{\rm s}(x)$ respectively by \citet{ful17}
and given by
[see equations (B2) and (B3) of the latter reference]
\footnote{
Equations (D2) and (D3) in \citet{ful17} contain typographical errors;
their right-hand sides are missing factors $\delta+1$ and $\alpha$ 
in the numerators respectively.
}
\begin{eqnarray}
 p_{\rm e}(x)
 & = &
 \frac{\lambda(x) p(x)}
  {\langle \lambda \rangle}
\label{eq:corappa5}
 \\
 & = &
 C \langle \lambda \rangle^{-1} \alpha x^{\delta+1} (1-x)^{\alpha-1}
\label{eq:corappa6}
\end{eqnarray}
and
\begin{eqnarray}
 p_{\rm s}(x)
 & = &
 \frac{1}{\langle \lambda \rangle}
 \int_x^1 dy \, \lambda(y) \eta(x|y) p(y)
\label{eq:corappa7}
 \\
 & = &
 C \langle \lambda \rangle^{-1} (\delta+1) x^\delta (1-x)^\alpha
\label{eq:corappa8}
\end{eqnarray}
with
\begin{eqnarray}
 \langle \lambda \rangle
 & = &
 \int_0^1 dx \, \lambda(x) p(x)
\label{eq:corappa9}
 \\
 & = & 
 \alpha + \delta + 2~.
\label{eq:corappa10}
\end{eqnarray}

The PDFs of the observable waiting times and sizes follow directly
from (\ref{eq:corappa5})--(\ref{eq:corappa10}).
The waiting time leading up to a glitch is the random value of 
$\Delta t$ generated by a Poisson process, 
whose rate $\lambda[x(t)]$ since the previous glitch
evolves deterministically due to spin down, 
conditional on the stress immediately after the previous glitch. 
The size of a glitch is the random value of $\Delta x=y-x$ 
generated by $\eta(x|y)$,
conditional on the stress $y$ immediately before the glitch.
Hence, applying equations (34) and (35) in \citet{ful17}, we obtain
\begin{eqnarray}
 p(\Delta t) 
 & = &
 \int_0^{1-\Delta t}
 dy \,
 p_s(y) p(\Delta t | y)
\label{eq:corappa11a}
 \\
 & = &
 (\alpha + \delta + 1) (1-\Delta t)^{\alpha+\delta}~,
\label{eq:corappa11b}
\end{eqnarray}
with $p(\Delta t | y)$ given by (\ref{eq:cor4}),
as well as
\begin{eqnarray}
 p(\Delta x) 
 & = & 
 \int_{\Delta x}^1 
 dy\, p_{\rm e}(y) \eta(y-\Delta x | y)
\label{eq:corappa12a}
 \\
 & = &
 (\alpha + \delta + 1) (1-\Delta x)^{\alpha+\delta}~.
\label{eq:corappa12b}
\end{eqnarray}
The PDFs (\ref{eq:corappa11b}) and (\ref{eq:corappa12b})
qualitatively resemble those observed in the pulsars
in Table \ref{tab:cor1} but they do not match the data in detail,
because the separable form of $\eta(x|y)$ in (\ref{eq:corappa2}) 
represents an approximation.
The moments, however, and their scalings with $\alpha$
are insensitive to the functional form of $\eta(x|y)$.
In particular,
the first moment of $p(\Delta t)$ evaluates to yield the important result
\begin{equation}
 \langle \Delta t \rangle 
 = 
 (\alpha + \delta + 2)^{-1}~,
\label{eq:corappa13}
\end{equation}
which is used heavily in \S\ref{sec:cor4};
see also Appendix A in \citet{ful17}.

\subsection{Size-waiting-time correlations
 \label{sec:corappab}}
To calculate the correlation coefficients $r_\pm$, 
we must first evaluate the joint probability of measuring
size-waiting-time pairs $(\Delta x,\Delta t_\pm)$.
There are subtleties involved.
Consider an arbitrarily selected sequence of three consecutive glitches
labelled by $G_1$, $G_2$, and $G_3$.
Suppose that $G_2$ has size $\Delta x$ and 
forward and backward waiting times $\Delta t_+$ and $\Delta t_-$ respectively.
Let $y_{\rm s}$ be the stress immediately after $G_1$.
Then deterministic evolution during the interval $G_1 G_2$
implies that the stress immediately before $G_2$ is 
$y_{\rm e} = y_{\rm s}+\Delta t_-$;
the event $G_2$ reduces the stress to 
$y_{\rm e} - \Delta x$
immediately after $G_2$;
and deterministic evolution during the interval $G_2 G_3$
implies that the stress immediately before $G_3$ is
$y_{\rm e} - \Delta x + \Delta t_+$.
Putting everything together,
the probability density of simultaneously measuring $\Delta x$
and $\Delta t_-$ given $y_s$ equals the conditional joint PDF
\begin{eqnarray}
 q_-(\Delta x,\Delta t_- | y_{\rm s})
 & = &
 p(\Delta t_- | y_{\rm s}) 
 \eta(y_{\rm s}+\Delta t_- - \Delta x | y_{\rm s}+\Delta t_-)~,
\label{eq:corappa14a}
 \\
 & \propto &
 (1-y_{\rm s})^{-\alpha}
 (1-y_{\rm s} - \Delta t_-)^{\alpha-1}
 \nonumber \\ 
 & & 
 \times
 (y_{\rm s} + \Delta t_-)^{-(\delta+1)}
 (y_{\rm s}+\Delta t_- - \Delta x)^\delta~,
\label{eq:corappa14b}
\end{eqnarray}
where $p(\Delta t_-|y_{\rm s})$ is given by (\ref{eq:cor4}).
Likewise, 
the probability density of simultaneously measuring $\Delta x$
and $\Delta t_+$ given $y_{\rm e}$ equals the conditional joint PDF
\begin{eqnarray}
 q_+(\Delta x,\Delta t_+ | y_{\rm e})
 & = &
 p(\Delta t_+ | y_{\rm e} - \Delta x) 
 \eta (y_{\rm e} - \Delta x | y_{\rm e})~,
\label{eq:corappa15a}
 \\
 & \propto &
 (1-y_{\rm e}+\Delta x)^{-\alpha}
 (1-y_{\rm e}+\Delta x - \Delta t_+)^{\alpha-1}
 \nonumber \\
 & &
 \times
 y_{\rm e}^{-(\delta+1)}
 (y_{\rm e}-\Delta x)^\delta~.
\label{eq:corappa15b}
\end{eqnarray}
where the first factor on the right-hand side of (\ref{eq:corappa15a})
is given again by (\ref{eq:cor4}).
The conditional joint PDFs are normalized according to
\begin{equation}
 1 =
 \int_0^{1-y_{\rm s}} d(\Delta t_-)
 \int_0^{y_{\rm s}+\Delta t_-} d(\Delta x) \,
 q_-(\Delta x,\Delta t_- | y_{\rm s})
\label{eq:corappa16a}
\end{equation}
and
\begin{equation}
 1 =
 \int_0^{y_{\rm e}} d(\Delta x)
 \int_0^{1-y_{\rm e}+\Delta x} d(\Delta t_+) \,
 q_+(\Delta x,\Delta t_+ | y_{\rm e})~.
\label{eq:corappa16b}
\end{equation}
The terminals on (\ref{eq:corappa16a}) and (\ref{eq:corappa16b}) 
ensure that the stress always stays in the domain $[0,1]$.

The law of total covariance states
\begin{equation}
 {\rm cov}(\Delta x,\Delta t_-)
 =
 {\rm E}[{\rm cov}(\Delta x,\Delta t_- | y_{\rm s})]
 +
 {\rm cov}[ {\rm E}(\Delta x|y_{\rm s}), {\rm E}(\Delta t_- | y_{\rm s}) ]~,
\label{eq:corappa17}
\end{equation}
where 
${\rm E}(\dots)=\int dy_{\rm s} \, p_{\rm s}(y_{\rm s}) \times (\dots)$
denotes the expectation value when marginalizing over $y_{\rm s}$,
and ${\rm E}(\Delta x|y_{\rm s})$ and
${\rm E}(\Delta t_-|y_{\rm s})$ are random variables themselves.
An analogous result applies to ${\rm cov}(\Delta x,\Delta t_+)$,
except that one marginalizes over $y_{\rm e}$.
It turns out that the $y_{\rm s}$ integrals in (\ref{eq:corappa17})
can be done analytically, viz.
\begin{eqnarray}
 {\rm E}(\Delta x|y_{\rm s})
 & = &
 \int_0^{1-y_{\rm s}} d(\Delta t_-)
 \int_0^{y_{\rm s}+\Delta t_-} d(\Delta x) \,
 \Delta x \,
 q_-(\Delta x,\Delta t_- | y_{\rm s})
\label{eq:corappa18}
 \\
 & = &
 \frac{1+\alpha y_{\rm s}}
  {(\alpha+1)(\delta+2)}~,
\label{eq:corappa19}
\end{eqnarray}
\begin{eqnarray}
 {\rm E}(\Delta t_-|y_{\rm s})
 & = &
 \int_0^{1-y_{\rm s}} d(\Delta t_-)
 \int_0^{y_{\rm s}+\Delta t_-} d(\Delta x) \,
 \Delta t_- \,
 q_-(\Delta x,\Delta t_- | y_{\rm s})
\label{eq:corappa20}
 \\
 & = &
 \frac{1-y_{\rm s}}{\alpha+1}~,
\label{eq:corappa21}
\end{eqnarray}
and
\begin{eqnarray}
 {\rm cov}(\Delta x,\Delta t_-|y_{\rm s})
 & = &
 \int_0^{1-y_{\rm s}} d(\Delta t_-)
 \int_0^{y_{\rm s}+\Delta t_-} d(\Delta x) \,
 \nonumber \\
 & & 
 \times
 [ \Delta x \Delta t_-
  - {\rm E}(\Delta x|y_{\rm s}) {\rm E}(\Delta t_-|y_{\rm s}) ]
 q_-(\Delta x,\Delta t_- | y_{\rm s})
\label{eq:corappa22}
 \\
 & = &
 \frac{\alpha (1-y_{\rm s})^2}
  {(\alpha+1)^2 (\alpha+2) (\delta+2)}~.
\label{eq:corappa23}
\end{eqnarray}
Upon substituting (\ref{eq:corappa19}), (\ref{eq:corappa21}), 
and (\ref{eq:corappa23}) into (\ref{eq:corappa17}), we obtain
\begin{equation}
 {\rm cov}(\Delta x,\Delta t_-)
 =
 \frac{\alpha}
  {(\delta+2) (\alpha+\delta+2)^2 (\alpha+\delta+3)}~.
\label{eq:corappa24}
\end{equation}
Similarly the total variances evaluate to give
\begin{equation}
 {\rm var}(\Delta x)
 =
 \frac{\alpha+\delta +1}
  {(\alpha+\delta+2)^2 (\alpha+\delta+3)}
\label{eq:corappa25}
\end{equation}
and 
${\rm var}(\Delta t_-) = {\rm var}(\Delta x)$.
Hence from (\ref{eq:corappa24}) and (\ref{eq:corappa25}) we arrive at
\begin{equation}
 r_-
 =
 \frac{\alpha}{(\delta+2)(\alpha+\delta+1)}
\label{eq:corappa26}
\end{equation}
for the correlation between sizes and backward waiting times.
Equation (\ref{eq:corappa26}) exhibits the same behavior
seen in Monte Carlo simulations and plotted in Fig.\ \ref{fig:cor4}.
The correlation increases with $\alpha$
but it asymptotes to a value $r_- \rightarrow (\delta+2)^{-1} \leq 1/2$,
which decreases as $\delta$ increases,
i.e.\ as small avalanches are favored more heavily.

The Pearson coefficient $r_+$ for the correlation between sizes
and forward waiting times is hard to calculate analytically.
Instead one can evaluate the integrals in the counterpart of 
(\ref{eq:corappa17}) numerically if required.
The result exhibits the same behavior seen in Monte Carlo simulations 
and plotted in Fig.\ \ref{fig:cor4},
i.e.\ the correlation $r_+$ decreases with $\alpha$.
It turns out that the relevant integrals diverge for $\alpha < 0.5$
for the specific form of $\eta(x|y)$ given by (\ref{eq:corappa2}).
The divergence can be fixed by cutting off the domain of integration
at some physically appropriate scale,
in the same way that a Cauchy PDF (for example)
does not have a well-defined mean or variance,
unless a cut-off is introduced.

\end{document}